\documentclass[fleqn,usenatbib]{mnras}

\usepackage{newtxtext,newtxmath}

\usepackage[T1]{fontenc}
\usepackage{comment}

\newcommand{\lya}{\ensuremath{\rm{Ly\alpha}}}
\newcommand{\lyb}{\ensuremath{\rm{Ly\beta}}}
\newcommand{\ttot}{\ensuremath{\tau_{\rm tot}}}

\newcommand{\tlya}{\ensuremath{\tau_{\rm{Ly\alpha}}}} 
\newcommand{\tlyb}{\ensuremath{\tau_{\rm{Ly\beta}}}}
\newcommand{\tlyc}{\ensuremath{\tau_{\rm{\alpha + \beta}}}}
\newcommand{\tlls}{\ensuremath{\tau_{\rm LLS}}} 
\newcommand{\Dlls}{\ensuremath{D_{\rm LLS}}} 
\newcommand{\lambo}{\ensuremath{\lambda_{\rm obs}}}
\newcommand{\lamba}{\ensuremath{\lambda_{\alpha}}}
\newcommand{\lambb}{\ensuremath{\lambda_{\beta}}}
\newcommand{\NHI}{\ensuremath{N_{\rm HI}}}
\def\HI{\hbox{\rm H~$\scriptstyle\rm I$}}

\newcommand{\stn}{\ensuremath{\rm S/N}}

\usepackage{graphicx}	
\usepackage{amsmath}	

\title[]{A New Measurement of the Mean Transmitted Flux in the Lyman-$\alpha$ and Lyman-$\beta$ Forest}

\author[J. Ding et al.]{
Jiani Ding,$^{1,2}$\thanks{E-mail: jding@ucsc.edu / OrcID:0000-0003-4651-8510}
Piero Madau,$^{1,3}$\thanks{OrcID: 0000-0002-6336-3293}
J. Xavier Prochaska$^{1,4}$
\\
$^{1}$Department of Astronomy and Astrophysics, University of California, Santa Cruz, 1156 High Street, Santa Cruz, CA 95064, USA\\
$^{2}$Department of Astronomy, Tsinghua University, Beijing 100084, China\\
$^{3}$Dipartimento di Fisica ``G. Occhialini", Università degli Studi di Milano-Bicocca, Piazza della Scienza 3, I-20126 Milano, Italy\\
$^{4}$Kavli IPMU (WPI), UTIAS, The University of Tokyo, Kashiwa, Chiba 277-8583, Japan}

\pubyear{2015}

\begin{document}
\label{firstpage}
\pagerange{\pageref{firstpage}--\pageref{lastpage}}
\maketitle

\begin{abstract}
We present new measurements of the mean transmitted flux in the hydrogen \lya\ and a relative transmitted flux measurement in \lyb\ using 27,008 quasar spectra from the Fourteenth Data Release (DR14) of the Extended Baryon Oscillation Spectroscopic Survey
(eBOSS). Individual spectra are first combined into 16 composites with mean redshifts in the range of $2.8<z<4.9$. We then apply Markov Chain Monte Carlo (MCMC) inference to produce a piecewise fit of the effective \tlya\ (corrected for metal lines and optically thick absorption) assuming a spline point distribution. 
We also perform a relative $\Delta \tlyb$ measurement with the same data set, finding $\Delta \tlyb<0.35$ at $z<4.8$. The 6-8 $\%$ precision measurements 
in the rest frame 1075-1150 \AA\ at $\it{z} \ < \rm{4.0}$ and 10-12 $\%$ precision measurements in the same region at $\it{z} \ > \rm{4.0}$ on $\tlya$, and 
our determinations of $\Delta \tlyb$, are dominated by systematic errors, 
likely arising from bias and uncertainties in estimates of the quasar continuum.
Our \tlya\ values show a smooth increase by a factor of 5 over the redshift range $z=2.4-4.4$.
\end{abstract}

\section{INTRODUCTION} \label{sec:style}
Intergalactic hydrogen scatters Lyman series radiation and produces a ``forest" of absorption lines in the spectra of distant quasars. This forest is a powerful probe of the evolution of cosmic baryons and of the formation of large-scale structures in the Universe, as it traces gas density fluctuations, the underlying dark matter distribution, and the ionization state and temperature of the diffuse intergalactic medium (IGM) following reionization \citep[see, e.g.,][]{meiksin2009,mcquinn2016_rev,gnedin2022}.
The primary second-order statistic derived from spectroscopic data -- the 1D power spectrum of the flux distribution in the forest -- 
provides one of the best tools for measuring the smoothness of the density field and constrain the nature of the dark matter (e.g., ``cold" versus ``warm"), cosmological parameters, and the thermal and reionization history of the IGM \citep[e.g.,][]{Palanque-Delabrouille2013,viel2013,Palanque-Delabrouille2015,Nasir2016,Ir2017,Yeche2017,Chabanier2018,garzilli2019,villasenor2022}. 

While the flux power spectrum contains information encoded across  different spatial scales, a more basic quantity, the effective optical depth of the forest $- \ln \, {F}(z)$, where $F(z) \equiv \langle F^i(z)\rangle$ is the mean transmitted flux at a given redshift, gives insight into the state of diffuse baryons, yields a  global measurement of the overall \HI\ content of the highly ionized IGM, and allows for estimates of  the intensity of the ionizing background radiation produced by star-forming galaxies and active galactic nuclei 
\citep[AGN;][]{Bolton2005,Becker2013b}. 
%
%
%
Specifically, gas in the IGM comprises most of the baryons in the Universe and approximately follows a density–temperature power-law relation of the form \citep[e.g.][]{Hui1997,puchwein15, McQuinn2016,villasenor22},
\begin{equation}
    T=T_0\Delta^{\gamma-1}
\end{equation}
where $\Delta=\rho_b/{\bar \rho}_b$ is the gas over-density, 
$T_0$ is the temperature at the mean cosmic density ${\bar \rho}_b$, and $\gamma-1$ is the power-law index of the relation. The time evolution of the parameters $T_0$ and $\gamma$ is determined by photoheating from hydrogen and helium ionization, recombination, Compton, and expansion cooling,  and collisional processes.  Assuming the IGM is in photoionization equilibrium, the optical depth of Lyman series lines can be linked to these properties and thus can in principle constrain $T_0$, $\gamma$, and the photoionization rate per hydrogen atom $\Gamma_{\rm HI}$.

There are different ways to perform a  measurement of the mean \lya\ transmission through the forest. One is to estimate the unabsorbed continuum in either high or medium resolution spectra and measure the transmitted flux as
$F^i = f^{i}_{\rm obs}/f^{i}_{\rm cont}$, 
where $f^{i}_{\rm obs}$ and $f^{i}_{\rm cont}$ are the observed flux and the unabsorbed continuum at each pixel, respectively
\citep[e.g.][]{Songaila2004,Kirkman2005,Dall2008}. Perhaps the most accurate determinations with this technique are those of \citet{Faucher2008}, who measured the  \lya\  transmission in 86 high resolution quasar spectra.
The common challenge faced by these studies is the difficulty of  identifying the peaks of transmission near unity, as the finite resonant scattering opacity in even the most underdense regions of the IGM causes a systematic continuum bias that becomes more severe with increasing redshifts  \citep{Faucher2008}. This issue does not affect determinations based on extrapolating the continuum from redward of the \lya\ line, as usually done in the case of lower resolution data for which a direct local continuum estimation is not feasible \citep[e.g.][]{McDonald2005,paris2011}. 

Another approach is to perform relative measurements using medium resolution quasar spectra in large datasets such as the Sloan Digital Sky Survey (SDSS). Here one exploits the general lack of evolution in the mean unabsorbed quasar continuum to compute, without continuum fitting, the mean transmitted flux $F(z)$, as a fraction of its value at $z\le 2.5$, then converts to absolute values by scaling to the measurements made by \citet{Faucher2008} from high-resolution data, where continuum errors are minimal. Using this technique, \citet{Becker2013} presented a new and highly precise measurement of the mean transmitted \lya\ flux over $2<z<5$ using a sample of 6065 moderate-resolution quasar spectra drawn from SDSS DR7.
More recently, \citet{kamble2020} have used a similar modeling framework, except for allowing spectral diversity in the sample, 
to measure the effective optical depth in the \lya\ forest using 40,035 quasar spectra from SDSS DR12. The higher opacity estimated 
at $z\lesssim 3$ by these authors compared to \citet{Becker2013} appears to require a 25\% weaker ionizing background.

The release of DR14 \citep{Abolfathi2018} of the Extended Baryon Oscillation Spectroscopic Survey (eBOSS) provides us with a large statistical sample with which to improve over and extend previous results by incorporating the \lyb\ forest at rest-frame wavelengths 
between 978 and 1014 \AA\ \citep{Irsic2013}. Because of the smaller cross-section, the study of \lyb\ absorption allows a better measurement of the equation of state of the IGM at higher overdensities, can break degeneracies when used in conjunction with \lya\ absorption, and may yield stronger constraints on feedback processes and other nuances of IGM physics that affect higher density regions but leave low-density structures intact \citep{Dijkstra2004}.

In this Paper, we present a refined determination of the effective opacity, \tlya, of the \lya\ forest (excluding high column
density absorbers),  
and a new measurement of 
the relative effective opacity, $\Delta$\tlyb, of the \lyb\ forest, based on
$27,008$ quasar spectra at redshift 
$2.8 < z < 4.8$ from SDSS DR14. The plan is as follows. In \S{2} we describe the dataset and our method of measuring the \lya\ and relative \lyb\ 
opacities in the IGM. The main results of this project are presented in \S{3}. We discuss our findings and compare 
them to previous measurements in \S{4}.

%

\begin{table}
    \centering
    \caption{Composite Spectra. Columns give the number of quasars included in each composite, the minimum ($z_{\rm min}$) and maximum ($z_{\rm max}$) quasar redshifts in each bin, and the mean redshift ($z_c$) of the composite.}
    \label{tab:table1}
    \begin{tabular}{cccccc}
        \hline
        Number of QSOs & $z_{\rm min}$ & $z_{\rm max}$ & $z_c$ \\
        \hline
        4640 & 2.8 & 2.9 & 2.85 \\
        4330 & 2.9 & 3.0 & 2.95 \\
        3995 & 3.0 & 3.1 & 3.05 \\
        3556 & 3.1 & 3.2 & 3.15 \\
        2830 & 3.2 & 3.3 & 3.25 \\
        2102 & 3.3 & 3.4 & 3.35 \\
        1189 & 3.4 & 3.5 & 3.45 \\
        991  & 3.5 & 3.6 & 3.55 \\
        1100 & 3.6 & 3.7 & 3.65 \\
        855  & 3.7 & 3.8 & 3.75 \\
        616  & 3.8 & 3.9 & 3.85 \\
        360  & 3.9 & 4.0 & 3.95 \\
        259  & 4.0 & 4.2 & 4.08 \\
        101  & 4.2 & 4.4 & 4.29 \\
        52   & 4.4 & 4.6 & 4.49 \\
        32   & 4.6 & 4.8 & 4.69 \\
        \hline
    \end{tabular}
    \vspace{0.3cm}
\end{table}


\section{DATA SET AND METHOD} 

In this section we describe how we construct composite spectra from the eBOSS DR14 dataset, and measure the normalized transmitted flux in the Lyman series forest at redshift $z=\lambda_{\rm obs}/\lambda_n-1$, where $\lambda_{\rm obs}$ is the observed wavelength at each pixel and $\lambda_n=1215.67\,$\AA\ for \lya\ and {\bf $1025.72\,$\AA\ } for \lyb.  The final dataset consists of $27,008$ quasar spectra at $2.8<z<4.8$ covering the wavelength interval $3800-9200\,$\AA\ with a resolution ranging from $R=1560$ at $3700\,$\AA\ to $R=2270$ at $6000\,$\AA\ (blue spectrograph), and from $R=1850$ at $6000\,$\AA\ to $R = 2650$ at $9000\,$\AA\ (red spectrograph).
The average emission redshift of quasars in each of our 16 composite spectra and the number of spectra in each composite are listed in Table \ref{tab:table1}. 

\begin{figure*}
\centering
\vspace{-1.0cm}
\includegraphics[width=0.97\textwidth]{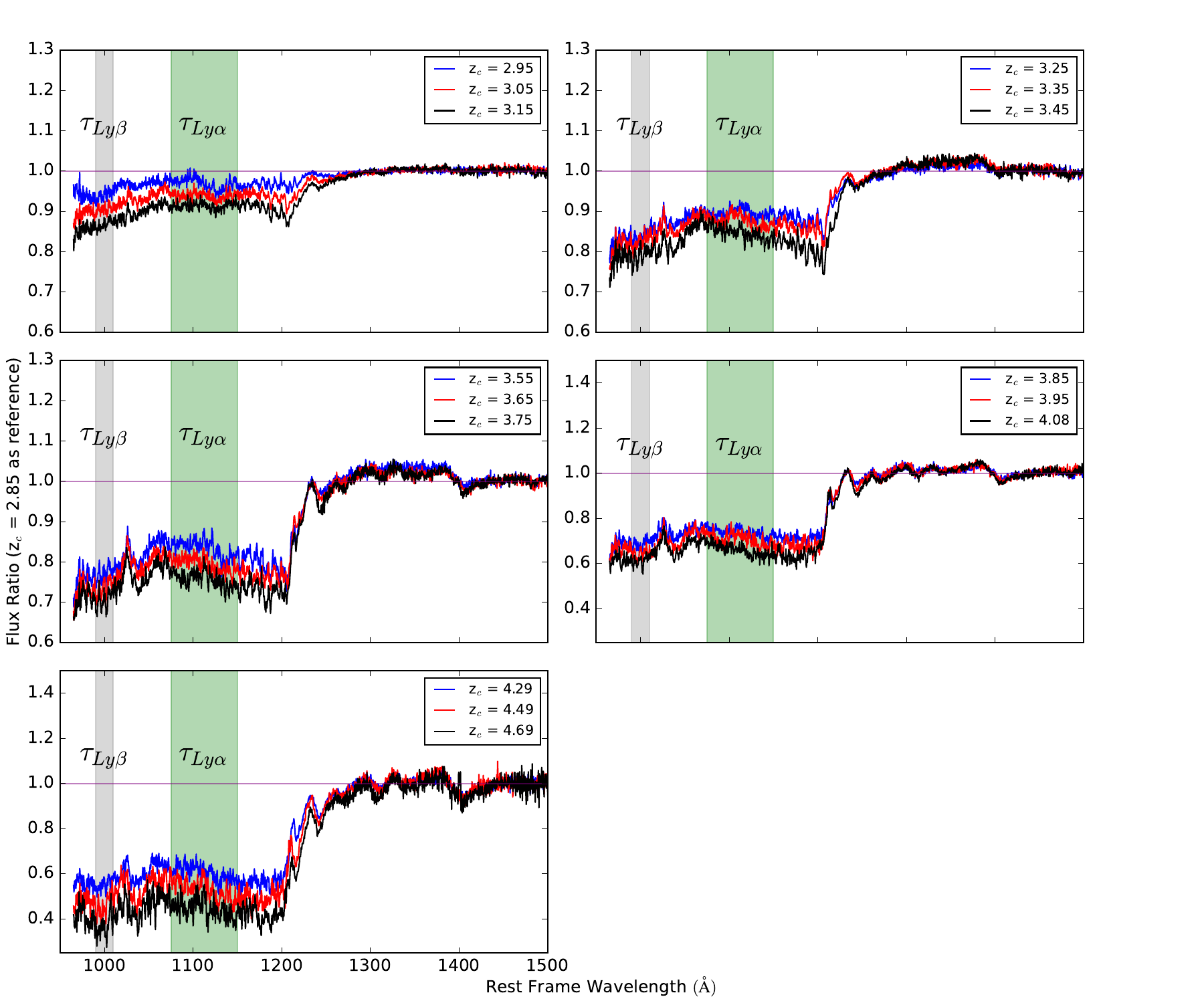}
\caption{The composite quasar spectra used in this work, plotted versus rest-frame wavelength. The grey region denotes the wavelength range used for \lyb\ measurement and the green region denotes the wavelength ranges used for \lya\ measurement. 
All spectra have been divided by our reference composite at $z_c=2.85$ and 
the horizontal line marks unity flux ratio.
Redward of \lya, 
all ratios approach one, as expected in the case of almost identical intrinsic quasar SEDs.
}
\label{fig:qso_sed}
\end{figure*}

Rather than fit continua to individual spectra, we shall assume below that the mean unabsorbed spectral energy distribution (SED) of quasars is very similar regardless of redshifts, and that the 
difference between composites at different epochs essentially reflects the evolution of the mean transmitted flux in the forest. The validity of this premise is illustrated in 
Figure \ref{fig:qso_sed}, where we plot the ratio between our composites and the reference composite spectrum at $z_c=2.85$. Longward of \lya, all ratios approach unity within the errors, as expected in the case of intrinsic SEDs that vary very little with redshift.
In contrast, shortward of rest wavelengths
$1215\,$\AA,
one observes strong and systematic evolution
in the flux ratio indicative of an increasing
IGM opacity with increasing redshift. When generating composite spectra in the quasar rest-frame, nearest-pixel values were adopted without interpolation (in order to reduce the covariance in the final composite), and combined by using an unweighted mean method
(all sightlines contributing equally). 
Individual spectra were normalized at the rest-frame wavelength $1440-1460\,$\AA\ where the quasar SED is relatively flat (without prominent emission and absorption lines).

\subsection{Systematic Errors}
\begin{figure*}
\centering
\vspace{-0.1cm}
\includegraphics[width=1.0\textwidth]{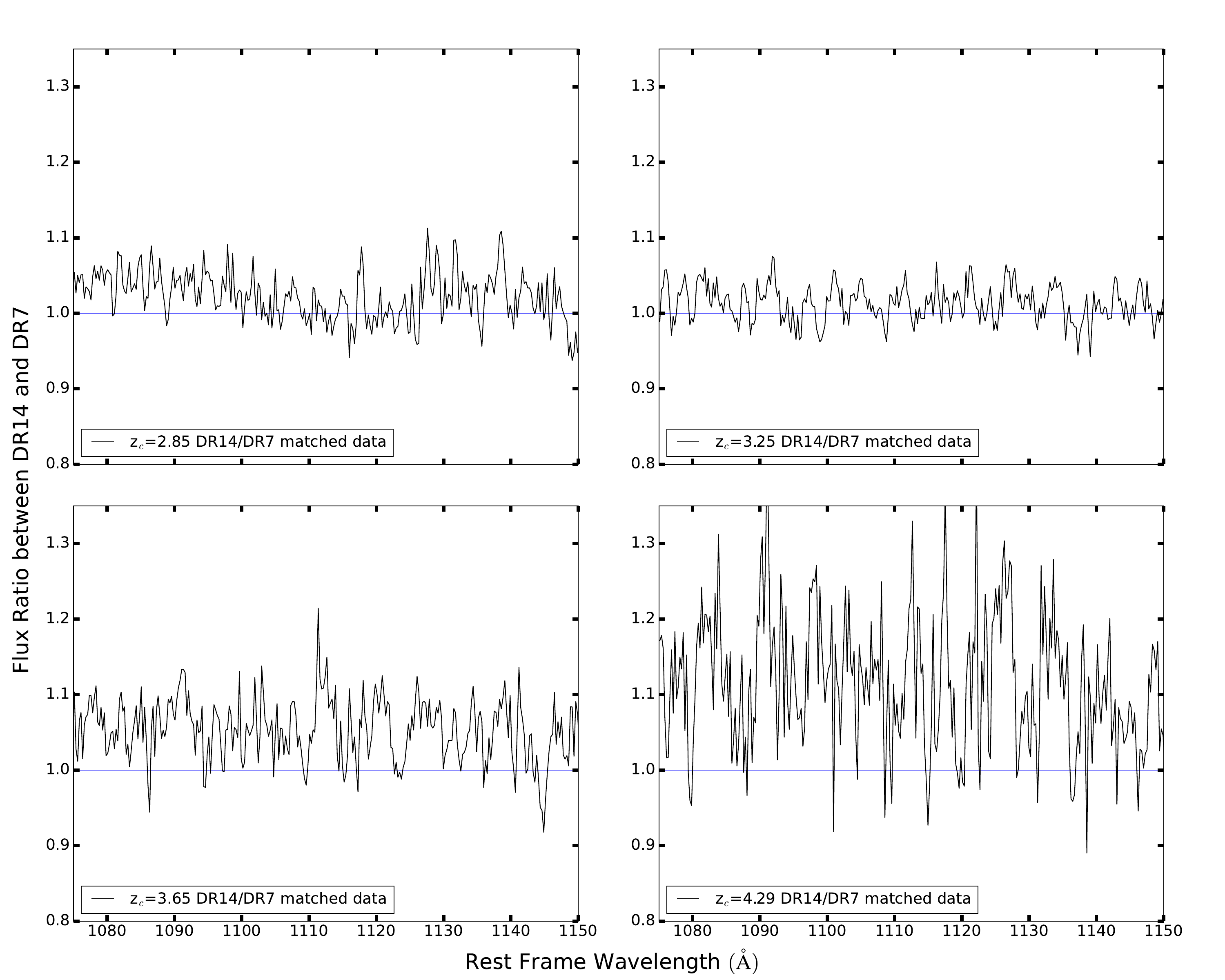}
\caption{Flux ratio of composite spectra (same quasar sample) from SDSS DR14 and DR7 at four different redshifts. Note how DR14 composites have a systematically higher flux level compared to DR7, a bias that increases with increasing redshift.
}
\label{fig:DR14_DR7}
\end{figure*}

The initial eSDSS DR14 spectra have systematic errors in the blue flux calibration and sky subtraction
\citep{lan2018}. Though corrected by the pipeline following \citet{Margala2016}, DR14 spectra at wavelength $\lambda_{\rm obs}<4000\,$\AA\ appear still slightly tilted compared to SDSS DR7 composites of the same quasar sample. This can be seen in 
Figure \ref{fig:DR14_DR7}, where the ratio 
between eBOSS DR14 and SDSS DR7 composites are plotted at four representative redshifts ($z_c=2.85,3.25,3.65,4.29$). 
The DR14 spectra have systematically higher fluxes (median ratios always greater than one) across the relevant redshift range. 
The comparison between DR14 and DR7 composites also suggests a systematic trend
of higher flux ratios towards increasing redshifts. We believe this bias is mainly due to  differences in the sky subtraction between the two data releases.

\begin{figure*}
\centering
\vspace{-1.0cm}
\includegraphics[width=1.0\textwidth]{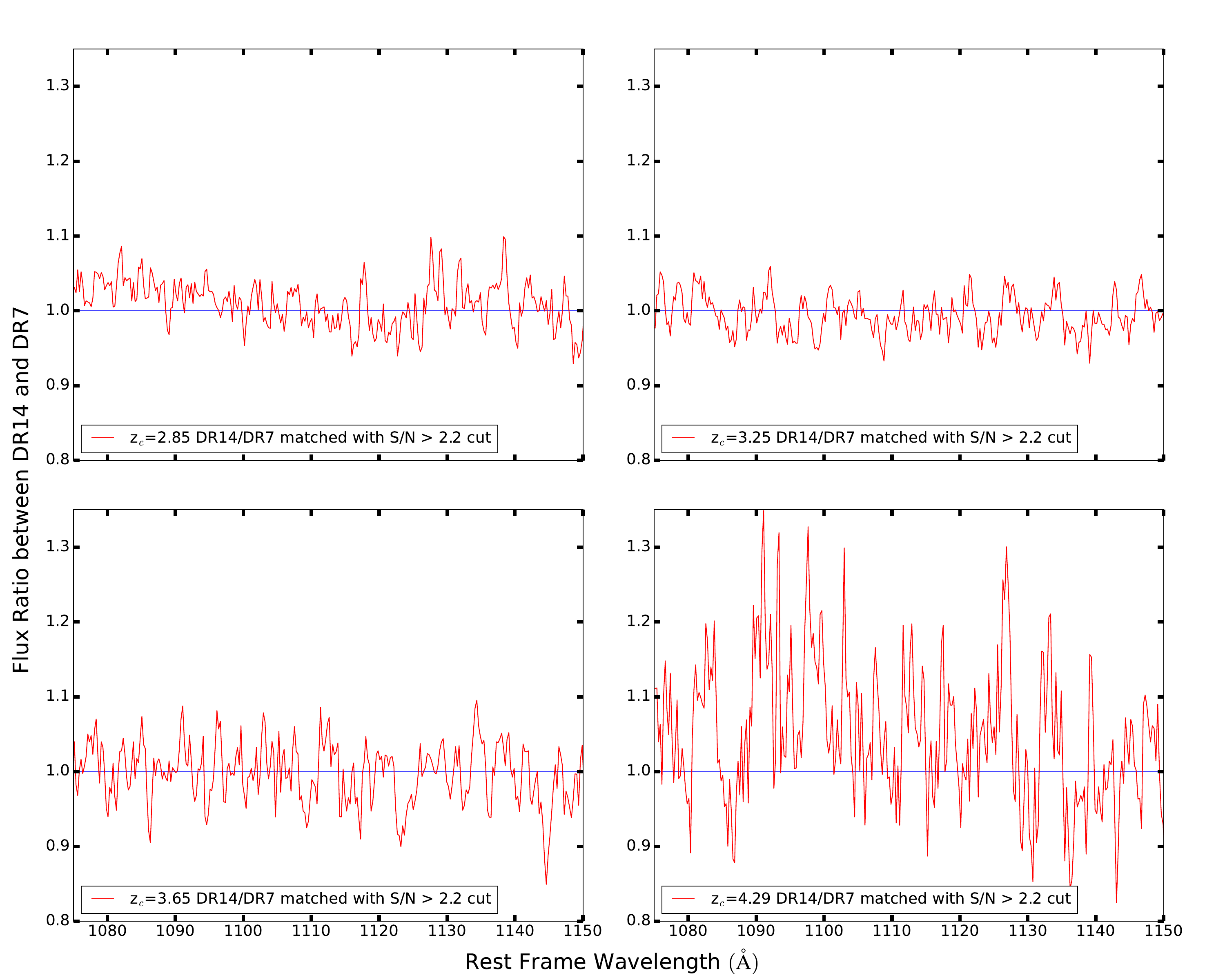}
\caption{Same as Fig. \ref{fig:DR14_DR7} but after a $\stn>2.2$ pixel$^{-1}$ cut.
}
\label{tab:samecompare_low}
\end{figure*}

In an attempt to reduce systematics, 
we have made different signal-to-noise (\stn)
cuts to our sample spectra and compared the resulting
DR14/DR7 composites in different redshift bins.
We find that a $\stn >2.2$ pixel$^{-1}$ 
cut at rest frame wavelengths of $1440-1460\,$\AA\ is optimal, as it greatly
reduces systematics while only reducing the
sample size by half. As shown in Figure  \ref{tab:samecompare_low}, with this \stn\ cut the DR14 and DR7 composites now match to within $3$\% in most redshift bins. 

\subsection{Final Composite Spectra}

\begin{figure*}
\centering
\vspace{-0.1cm}
\includegraphics[width=1.\textwidth]{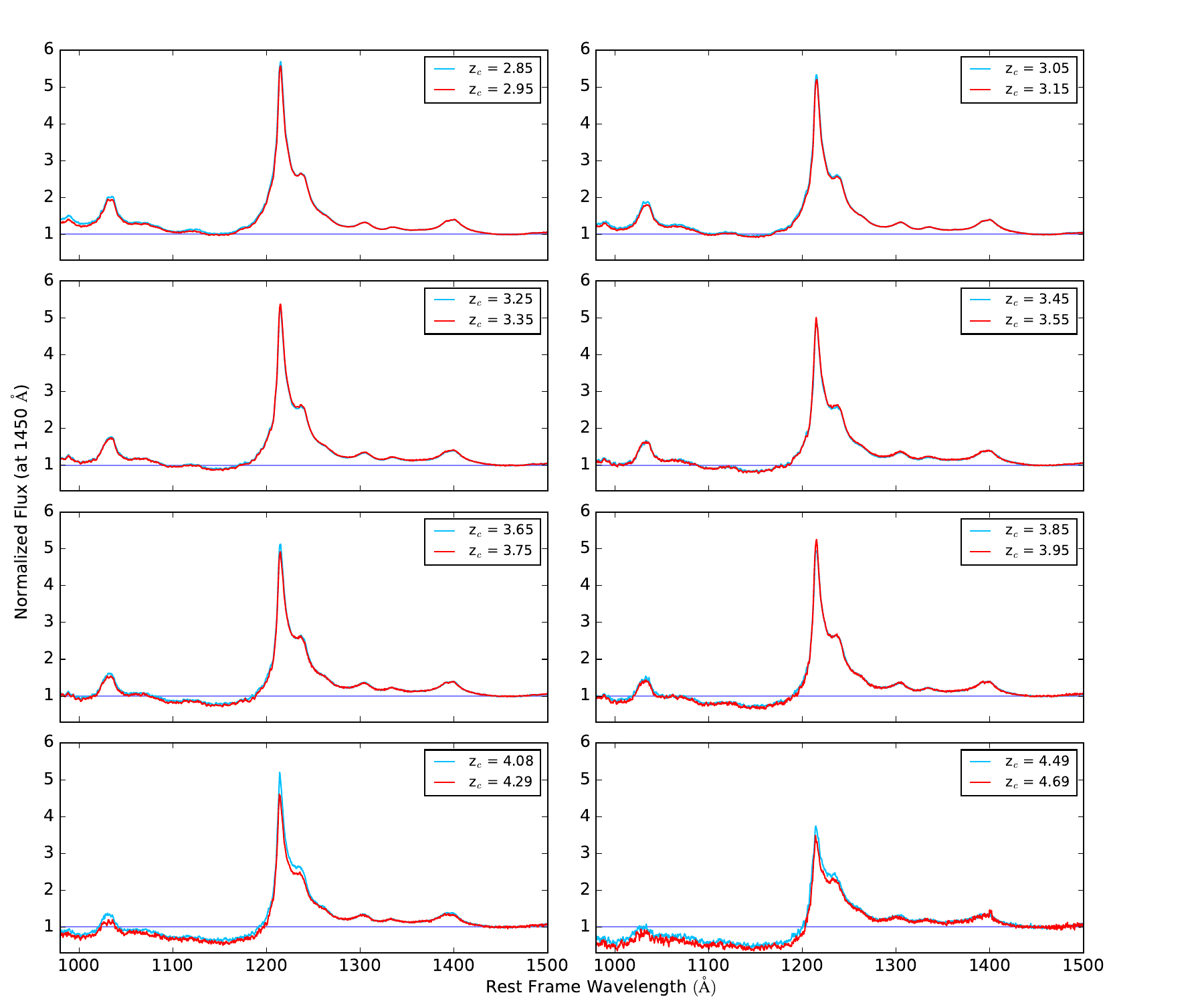}
\caption{Final composite spectra from eBOSS DR14 data in different redshift bins.
}
\label{fig:composite}
\end{figure*}

The final sample consists of $27,008$ individual quasar spectra, which are combined in 16 redshift bins of width equal to 0.1 for quasars with redshifts $<4.0$, and 0.2 at higher redshifts because of the decreasing number of targets.
We exclude spectra with broad absorption line (BAL) features using the relevant flag in the quasar catalog \citep{Paris2018},  and mask the skylines region from 5570 to 5590 \AA. The final composite spectra for eBOSS DR14 are shown in Figure \ref{fig:composite}.

\section{Mean Transmitted Flux}

\subsection{\lya\ Opacity} 

The mean transmitted flux at the rest-frame wavelength $\lambda_c$ of a quasar composite with mean emission redshift $z_c$ can be written as 
\begin{equation}
    F(z) = {f_{\rm obs}(\lambda_c,z_c)\over
    f_{\rm cont}(\lambda_c)},
\end{equation}
where, in the \lya\ forest, $1+z_{abs}=\lambda_c(1+z_c)/\lambda_\alpha$.  \citet{Becker2013} measured the ratio of $F(z)$ at 
two different redshifts $z_1$ and $z_2$ by evaluating
the ratio of fluxes between composites ``$c,1$'' and ``$c,2$'' at the same rest-frame wavelength,
\begin{equation}
    {F(z_1)\over F(z_2)} = {f_{\rm obs}(\lambda_c,z_{c,1})\over
    f_{\rm obs}(\lambda_c,z_{c,2})},
\end{equation}
a relation that
is satisfied  when the mean continuum in the composites either does not evolve or has been appropriately corrected vs. redshift. 
They compute $F(z)$ up to a normalization factor, achieved by fitting a function that matches the observed flux ratios. The flux $F(z)$ is parameterized discretely in bins of $z = 0.1$, with the initial bin at $z = 2.15$, and computed as a fraction of the transmitted flux in that bin, i.e., $F(z)/F(z_c=2.15)$.  

Here,  we adopt a modified approach. We assume, as in \citet{Becker2013}, that the quasar unabsorbed SED does not change with redshift (see Fig. \ref{fig:qso_sed}), but  use Monte Carlo Markov Chain (MCMC) inference to perform a simultaneous fit to all of our DR14 composite spectra using a two component model: 1) the continuum represented by a cubic spline that varies with rest-frame wavelength; and 
2) $\ttot(z)\equiv - \ln \, {F}(z)$, also modeled as a cubic spline as a function of redshift. 
We emphasize that \ttot\ includes
all sources of opacity in the \lya\ forest region including metals and optically-thick absorbers (discussed below).

Our empirical model for the shape of the quasar continuum uses 12~spline points spaced in wavelength as detailed in the Appendix (\ref{tab:location2}). The cubic-spline form for \ttot\ differs from  the power-law function commonly adopted in the literature (but shown to be insufficient by \citealt{Becker2013}). After experimentation, we chose ten spline points to represent the total
\lya\ effective opacity as a function of redshift. The prior uniform distribution of the first spline point ranges from 0.2 to 0.8, and the range limits for the next points increase by 0.045 (i.e. $0.245-0.845$ for the second point, $0.29-0.89$ for the third, etc (see Table~\ref{tab:location1}). 
Lastly, to yield an absolute estimate of \ttot, 
the model is constrained by the
high-resolution measurements of \cite{Faucher2008} 
at redshifts $2.0-2.8$, including our estimate for
optically-thick absorbers ($\S$~\ref{sec:LLS}) which
were masked by \cite{Faucher2008}.
We also note that this is a small correction ($\approx 0.015$)
at these redshifts.

\begin{figure*}
\centering
\vspace{-0.1cm}
\includegraphics[width=160mm]{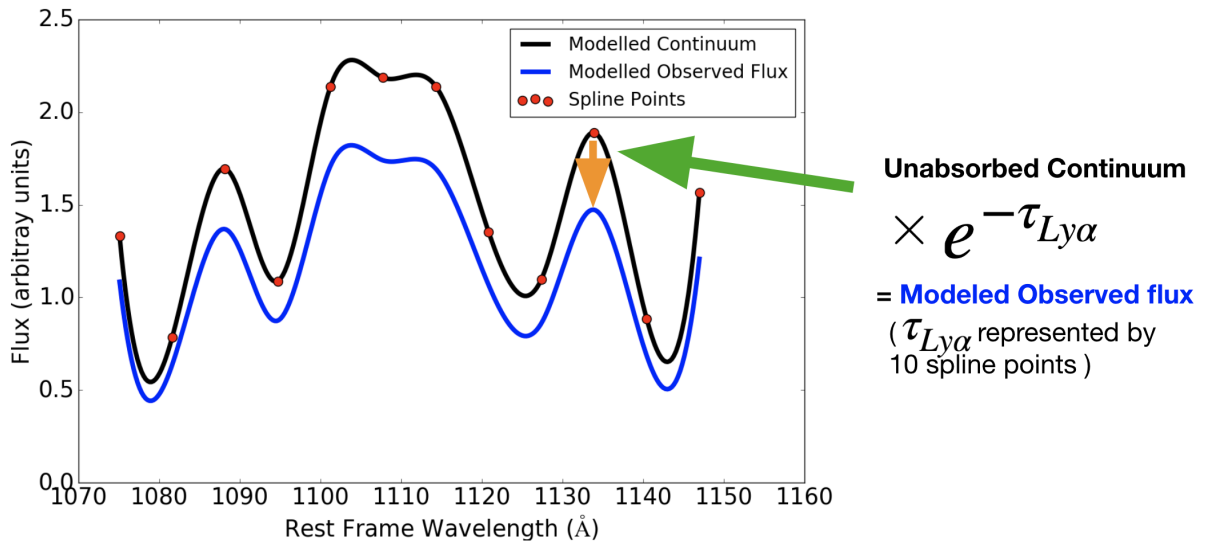}
\caption{Our MCMC method. {Red dots:} spline points used to represent the {unabsorbed quasar continuum}.  The mock observed flux (solid blue curve) is 
given by the unabsorbed continuum $\times \exp (-\tlya)$, where the function $\tlya(z)$ is represented with 10 spline points. The mock flux is then fitted to the DR14 composite spectra.
}
\label{fig:method}
\end{figure*}

Our Bayesian  MCMC analysis centers on the rest-frame wavelength region $1075-1150$ \AA\ in the composite spectra for two reasons: 1) the ionization state of 
intergalactic gas is known to be enhanced near quasars  \citep{Bajtlik1988,Lidz2007}, and restricting to wavelengths
$\le 1150$ \AA\  mitigates the impact of this ``quasar proximity effect"; 
and 2) the lower bound avoids variation in the quasar continuum from OVI and FeII emission. 
The prior distributions of each of the 22  MCMC parameters are assumed to be uniform, and their ranges are listed in Table \ref{tab:prior} of the Appendix. 
The fitting algorithm adopts a  multivariate normal log-likelihood distribution, with a covariance matrix generated as  described in the Appendix. Our MCMC method is illustrated in Figure \ref{fig:method}.

\begin{figure*}
\centering
\vspace{-0.1cm}
\includegraphics[width=130mm]{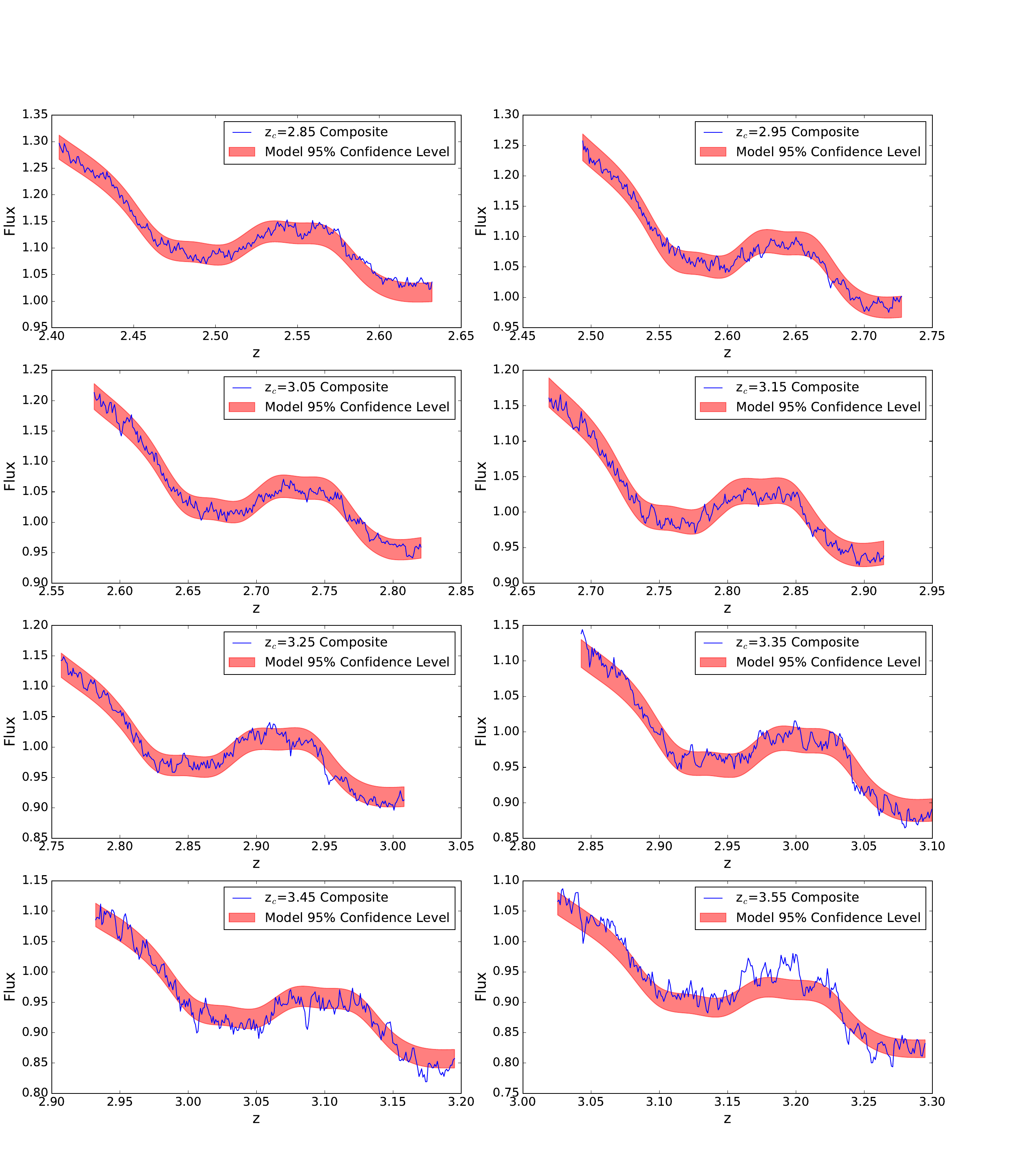}
\caption{MCMC best-fit model to DR14 composite spectra (blue curves) in the redshift interval $z_c=2.85-3.55$. The shaded region corresponds to the 95\% confidence range of the transmitted flux marginalized over the posterior distribution. 
}
\label{tab:modelfit1}
\end{figure*}

\begin{figure*}
\centering
\vspace{-0.1cm}
\includegraphics[width=130mm]{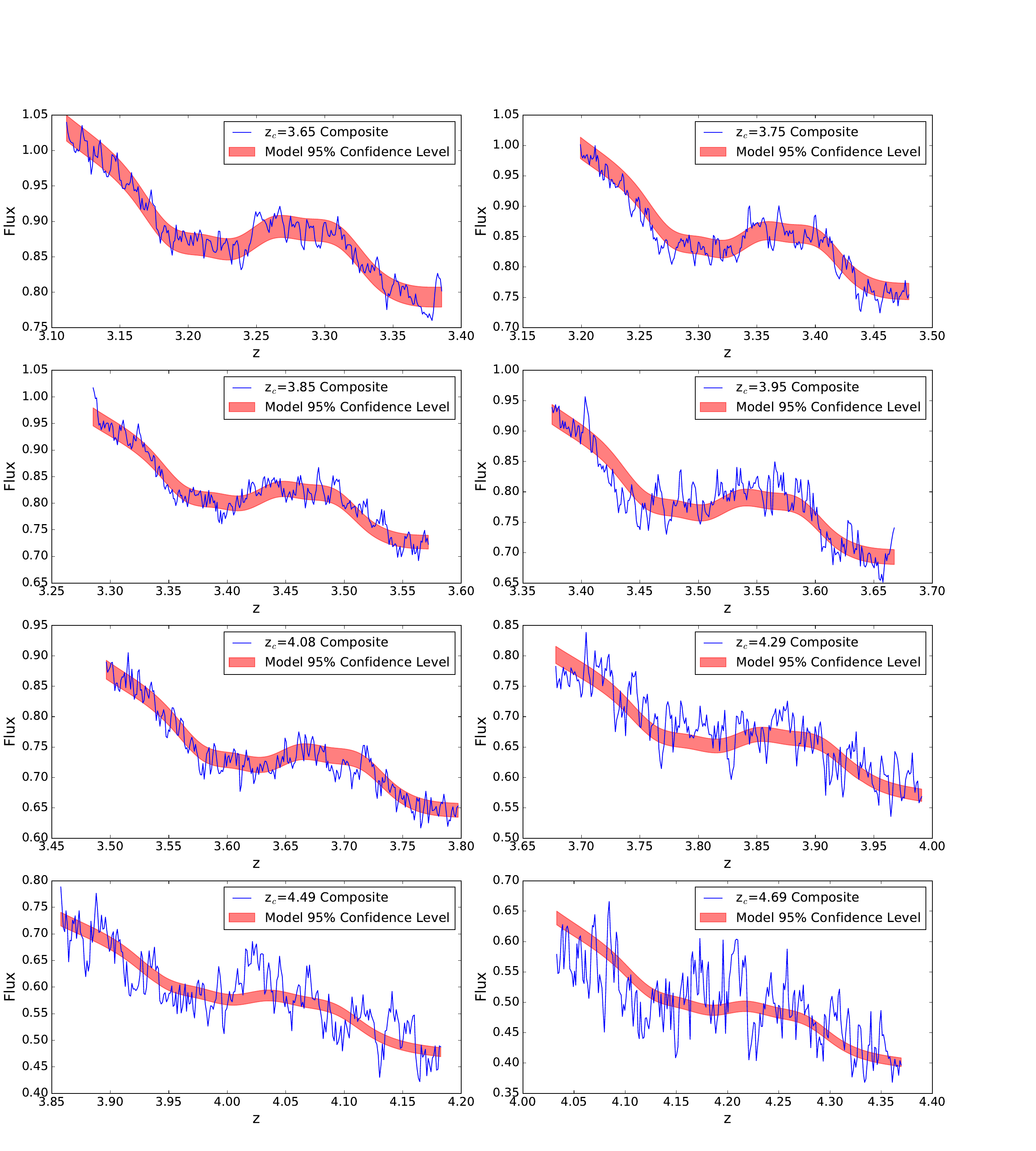}
\caption{Same as the previous figure in the redshift interval $z_c=3.65-4.69$.
}
\label{tab:modelfit2}
\end{figure*}

In the Appendix we show trace and kernel density plots for all the 22 spline parameters describing the quasar continuum (Fig. \ref{tab:taualpha1}) and $\tlya$ (Fig.  \ref{tab:taualpha2}). All parameters appear to have converged at the end of the MCMC sampling. 
Also in the Appendix, Figure \ref{fig:cornerplot1} 
depicts  the one- and two-dimensional projections of the posterior distributions for the quasar continuum.
The best-fit MCMC model is compared to SDSS DR14 composite spectra as a function of redshift in Figures \ref{tab:modelfit1} and \ref{tab:modelfit2}.

\subsection{Metal Line Contamination}
\label{sec:metals}

Metal absorption in the \lya\ forest
is an important systematic component to be modeled for a precise measurement of \tlya.
Previous studies have shown that the flux decrement associated with the opacity of metal lines is nearly independent of redshift \citep{Schaye2003,Kirkman2005, Becker2011}. 
Therefore, it presents only an arbitrary constant that we
choose not to model.  By using the metal-corrected 
\tlya\ measurements of \citet{Faucher2008} at $z = 2.0-2.8$ to constrain our MCMC model combined with optically-thick absorbers, 
we effectively recover an estimate without metals absorption.

\subsection{Correction for Optically Thick Absorbers}
\label{sec:LLS}

Our \lya\ opacity measurements (corrected for metal absorption) include all diffuse intergalactic gas along the line of sight including optically-thick absorbers with $\NHI \ge \ 10^{17.2} \ \rm{cm}^{-2}$.
These ``Lyman-limit systems" (LLS) arise from highly overdense regions in the circumgalactic environment and are difficult to model  analytically or in numerical simulations. It is standard practice, therefore, to subtract the  contribution of LLS to \tlya. Here, we follow \citet{Becker2013}, and compute the 
integrated flux decrement from LLS as
\begin{equation}\
    \Dlls=\frac{1+z}{\lambda_{\alpha}}{\int dN_{\rm{HI}}}\int db f(N_{\rm HI},z)W_{0}(N_{\rm{HI}},b),
    \label{eq:opticalthick}
\end{equation}
where 
$f(N_{\rm HI},z)$ is the neutral hydrogen column density distribution, and $W_{0}$ is the rest-frame equivalent width. We use the $f(N_{\rm HI},z)$ distribution for $3.0 < z < 4.5$ simulated by the package $\it{pyigm}$ 
\citep{Prochaska2014}, and 
integrate Equation (\ref{eq:opticalthick}) over the
column density range $\NHI = 10^{17.2}-10^{22}$ cm$^{-2}$. 
We assume that gas column density and Doppler parameters $b$ are statistically independent, and adopt a fixed value  of $b=24\,$ km s$^{-1}$ (varying $b$ in the range 
$20-30$ km s$^{-1}$ has little impact on our results). 
The opacity from these optically-thick absorbers
is therefore 
$\tlls \equiv -\ln(1-\Dlls)$.

\begin{figure*}
\centering
\vspace{-0.1cm}
\includegraphics[width=0.49\textwidth]{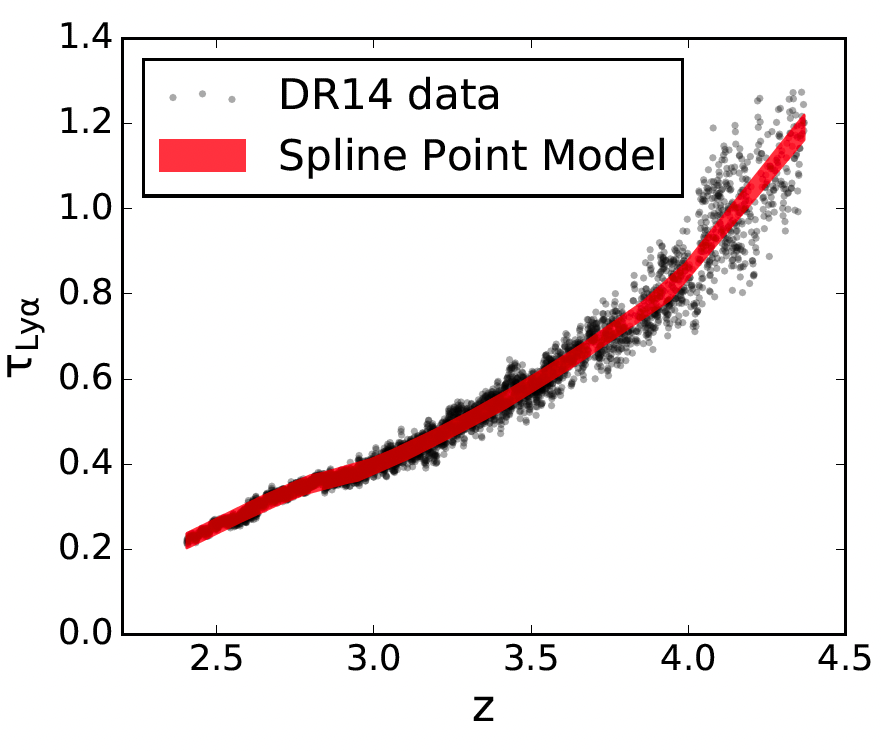}
\vspace{-0.5cm}
\caption{Observational determinations from DR14 composite spectra of the \lya\ opacity in the redshift range $2.4-4.4$. The best-fit from MCMC inference (corrected for metal line and LLS absorption) is shown as the 
shaded region corresponding to the 95\% confidence interval of \tlya\ marginalized over the posterior distribution.}
\label{fig:comparison_2}
\end{figure*}

In our analysis, we first derive a raw measurement of $\ttot'$ from MCMC inference, and then subtract the contribution of optically thick absorbers,
\begin{equation}
\tlya = \ttot' - \tlls
\end{equation}
where $\ttot'$ has been corrected for metal-absorption. The magnitude of the LLS correction is approximately 4-6$\%$. The fully corrected \lya\ opacity of the IGM is  shown in Figure \ref{fig:comparison_2}. Our flexible spline point model for \ttot\ reproduces the redshift evolution of DR14 composite spectra more precisely than the standard power-law approximation.

\subsection{\lyb\ Relative Opacity} 
\begin{figure*}
\centering
\vspace{-1.0cm}
\includegraphics[width=0.55\textwidth]{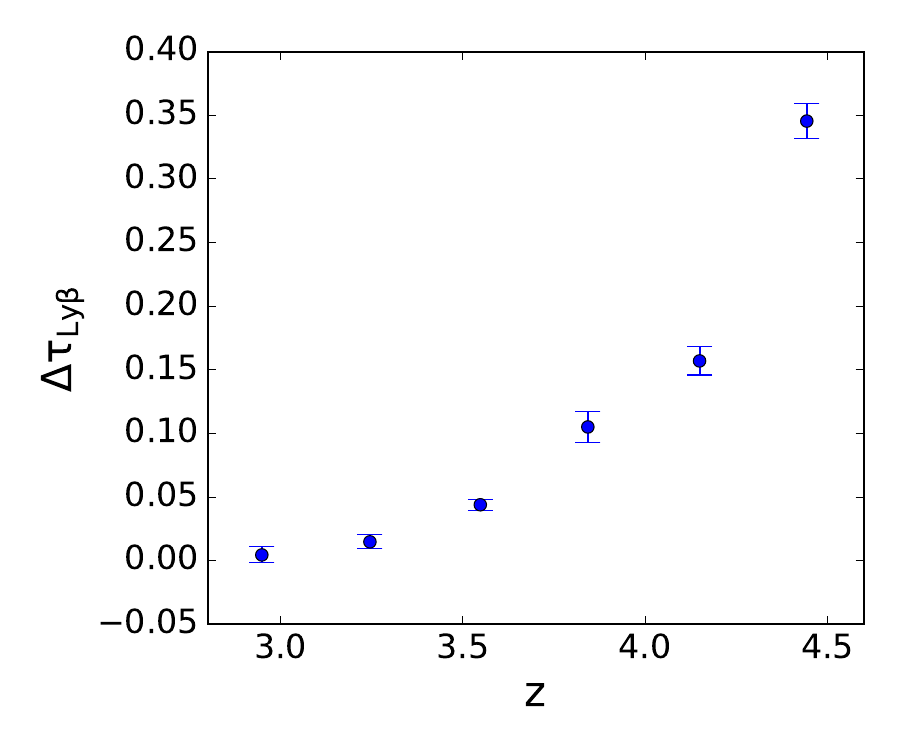}
\caption{Relative \lyb\ effective opacity of the IGM from eBOSS DR14 data. The ten redshift bins cover the range $2.8 \leq z \leq 4.6$.
}
\label{tab:beta}
\end{figure*}

Leveraging the large dataset of eBOSS DR14 and its blue wavelength coverage, we extend our analysis of the opacity of intergalactic hydrogen to include \tlyb. As there are no existing measurements of quasar UV continuum levels in the \lyb\ forest, we can only 
provide here a relative determination, $\Delta\tlyb$.
We focus on the rest-wavelength range $990-1010\,$\AA\ in the composite spectra, the region  
blueward of the \lyb\ emission line
where the signal-to-noise is higher.
As before, we measure the ratio of transmitted fluxes between all other composites and that at the reference redshift $z_c = 2.95$. In each composite, the transmission through the \lyb\ forest is determined by the combined effective
optical depth,
\begin{equation}
\tlyc(\lambo) = \tlya(z_\alpha) + \tlyb(z_\beta),
\end{equation}
for the overlapping spectral region, with
$z_\alpha = \lambo/\lamba -1$
and $z_\beta = \lambo/\lambb -1$.
Therefore, the flux ratio between a pair of composite spectra is sensitive to the difference 
$\Delta \tlya + \Delta \tlyb$ evaluated at \lambo. We first calculate \tlyc, then attain a value for $\Delta \tlya + \Delta \tlyb$ 
 evaluated at \lambo\ relative to $\tlyb(z_c=2.95)$ from equation (3).
 We subtract from it the foreground $\Delta\tlya$ and finally determine $\Delta\tlyb(z)$. 
 The evolving relative \lyb\ opacity in ten redshift bins is shown in  Figure \ref{tab:beta}. 

\section{DISCUSSION} 

\begin{figure*}
\centering
\vspace{0.0cm}
\includegraphics[width=150mm]{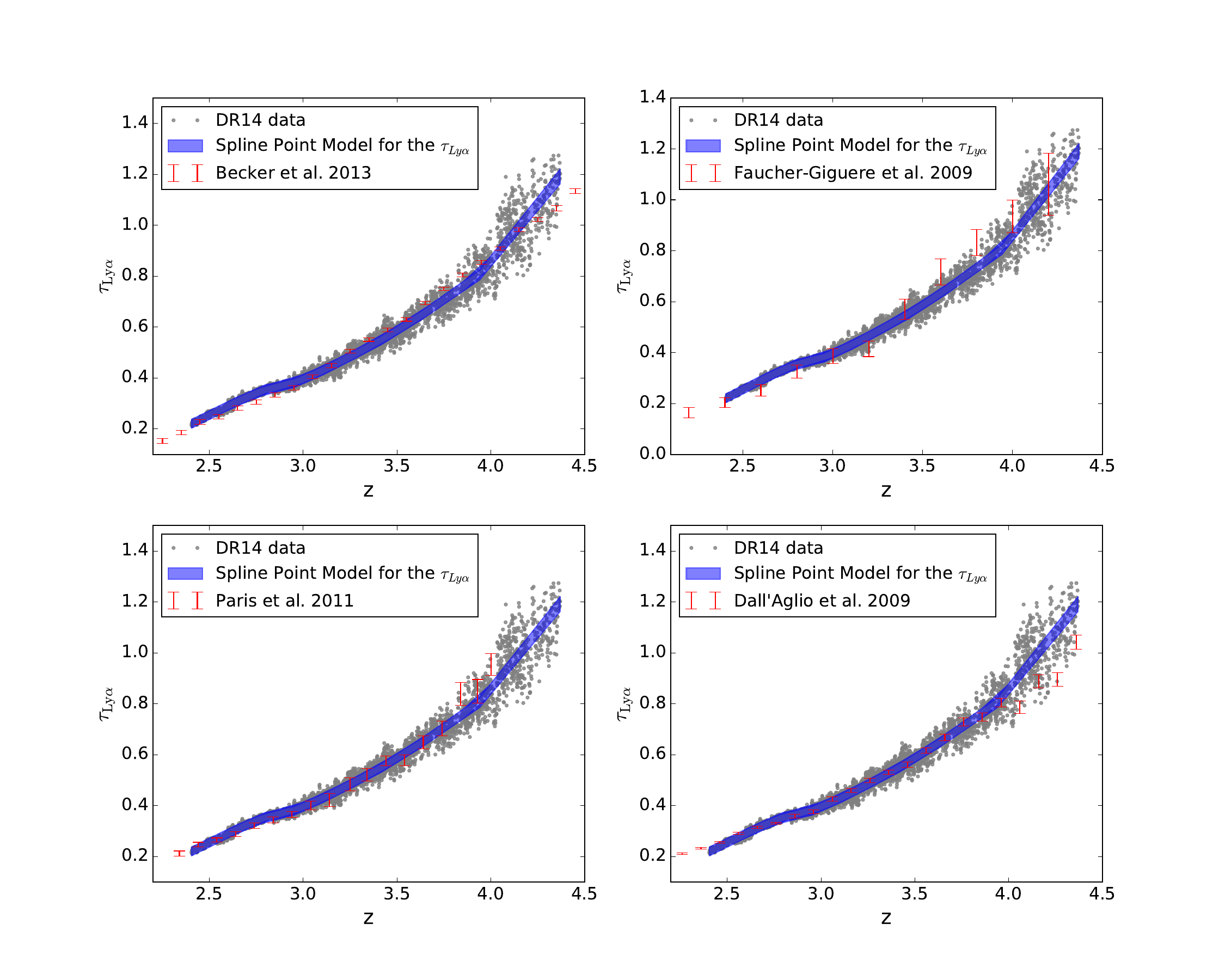}
\caption{The comparison between $\tlya$ measurement from the best fitted spline evolution for $\tlya$ (blue region marks 3-sigma confidence level for the fitted $\tlya$ model) from the MCMC modeling plotted with the DR14 data after correcting metal line absorption and optically thick absorbers and the measurement in \citet{Becker2013,Dall2009,Faucher2008,paris2011}. Upper left: The comparison between our results and measurement in \citet{Becker2013}. 
Upper right: The comparison between our results and measurement in \citet{Faucher2008}. Lower left: The comparison between our results and measurement in \citet{paris2011}. 
Lower right: The comparison between our results and measurement in \citet{Dall2009}. }
\label{fig:comparison_f}
\end{figure*}

We have presented new measurements of the mean transmitted UV flux through cosmic hydrogen using 27,008 quasar spectra from eBOSS DR14, and applied MCMC inference to produce a spline-piecewise fit of the effective  \lya\ (corrected for metal lines and hydrogen optically thick absorption) and relative \lyb\ optical depths. In Figure \ref{fig:comparison_f} we plot our derived intergalactic \lya\ opacity along with previous SDSS-based results by \citet{Becker2013}, \citet{Dall2009}, and \citet{paris2011}, as well as
with the high-resolution  determinations  made by \citet{Faucher2008} -- and used to constrain our MCMC model at $z<2.8$. 
All of the works yield similar \tlya\ estimations at $z<3$
and our new measurements lie between the locus of data at higher
redshifts.

\begin{figure*}
\centering
\vspace{-1.0cm}
\includegraphics[width=0.6\textwidth]{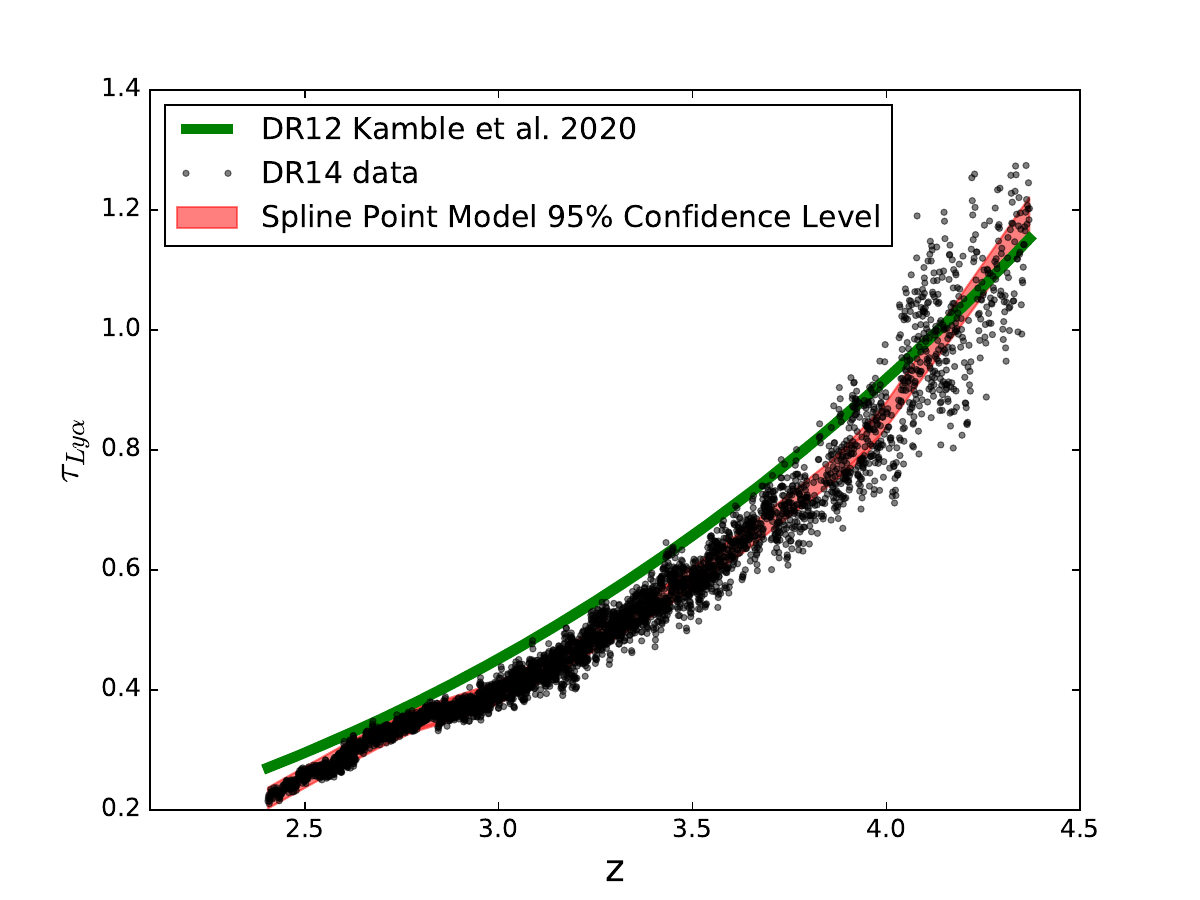}
\caption{Comparison between $\tlya$ measurement from the best fitted spline evolution for $\tlya$ (red region marks 3-sigma confidence level for the fitted $\tlya$ model) from the MCMC modeling plotted with the DR14 data after correcting metal line absorption and optically thick absorbers and the measurement in \citet{kamble2020} (green curve).}
\label{fig:comparison_k}
\end{figure*} 

Compared to \citet{Becker2013}, the $\approx 10\%$
differences between the two 
determinations at $z = 3.4 - 4$ may be associated with three different factors. Firstly, the difference in the
sky subtraction for the DR7 and DR14 data sets. 
However, because \cite{Becker2013} estimate that sky subtraction 
affects only $\approx 5\%$ of the results, we expect this
is a minor effect. 
Secondly, the DR14 sample analyzed here is considerably 
larger than the DR7 sample. Also, the systematic on the \tlya \ measurement caused by the Quasar SED shape discussed in \S APPENDIX D may contribute to $\approx 6-8\%$ of the measurement difference. 
Taking the three factors into account, our results are consistent with \cite{Becker2013}.
\citet{Dall2009} combined a redshift-dependent global correction from mock simulated spectra with a local spline interpolation to fit the continuum of 1733 individual quasar spectra. At the medium resolution of SDSS redshift-dependent corrections must be larger, and will be more significant at lower redshifts. There is a small discrepancy between \citet{Dall2009} measurement and our results at $z>$ 4.0. The reasons for this slight disagreement may be as follows. While fitting the continuum in the medium resolution spectra, the fitting parameters will also have larger error and systematics. Also, in \citet{Dall2009}, the mock spectra used for fitting the continuum are based on a fixed number density, column density and Doppler parameter distributions of absorbers. This also leads to the systematic effects causing some discrepancy in the evolution of the \lya\ opacity. \\
Finally, a more recent paper by \citet{kamble2020} utilizes BOSS DR12 data to measure the \lya\ opacity in a quasar sample that is similar to ours. These authors model \tlya\ as a power-law and fit for the continuum
normalization and then marginalize over the latter in seven bins
of quasar samples.
A comparison between their results and ours shows  (Fig. \ref{fig:comparison_k})  a systematically higher 
optical depth.  The majority of the difference originates 
from their decision not to constrain their model by high-resolution
estimates of \tlya\ \citep[e.g.][]{Faucher2008}.
We also note that while they ignored quasars with foreground
damped \lya\ systems, their \tlya\ do include opacity from LLS
at lower column densities.
We estimate, however, that this is a relatively small correction
($<10\%$). Also, our measurement shows a small dispersion with \citet{kamble2020} at higher redshift. This may due to the difference in our methods to attain the \tlya\ measurement. Our measurement use a spline point model for the \tlya\ measurement while \citet{kamble2020} assume a power law distribution for the \tlya\ , leading to the fact that the discrepancy tends to get small in the high redshift regime. In \S APPENDIX D, we also investigate the effect caused by different spectral indexes (similar to the spectral index groups in \cite{kamble2020}). Taking into account the systematic discussed in \S APPENDIX D, our discrepancies with \cite{kamble2020} are largely relieved. 

A recent publication from the large spectroscopic datasets (the Dark Energy Spectroscopic Instrument \citep{Turner2024} also perform a measurement on \tlya\ by using a Convolutional Neural Network to reconstruct the Lyman alpha continuum. Our results are consistent with the DESI results within 95$\%$ confidence level taking into account the systematic mentioned in \S APPENDIX D.

\section{conclusion}
We measure the \tlya\ and relative \tlyb\ using a novel method from the eboss DR 14 spectroscopic data. Our results are consistent with 
the majority of \tlya\ measurements in previous literature.
In the future, as we turn to the next generation of large spectroscopic datasets
\cite[the Dark Energy Spectroscopic Instrument; DESI/DESI-II][]{desi} ,
it is evident that quasar sample size is no longer the limiting
factor in estimating \tlya.  Instead, the accuracy and precision
is limited by systematic error in data collection (e.g.\ fluxing)
and the assumption that quasar continua are invariant with redshift.
Nevertheless, 
opportunities to extend to higher order Lyman series lines will
materialize, both in surveys like DESI but also in higher-resolution
samples.  

\section{Data Availability}
The data and algorithm used in this article are available in \href{https://github.com/JianiDing/sawtooth_project}{$sawtooth\_project$}, at \url{https://github.com/JianiDing/sawtooth_project}.

\section*{Acknowledgements}

Support for this work was provided by NASA through grants 80NSSC21K027 and 80NSSC22K0814
(P.M.). JD acknowledge support from the
National Science Foundation of China: 12073014 and China Manned Space Project: CMS-CSST2021-A05.

\bibliographystyle{mnras}
\bibliography{lyman_series_paper.bib}

\appendix

\section{Bayesian Inference Parameters and Priors}
We use a MCMC process to sample the best fitted parameters. All the priors for the 22 parameters are uniform. The location for the first 12 spline points (continuum model) is listed in \ref{tab:location2}. The location for the 10 spline points for the $\rm{\tau_{\alpha}}$ is listed in \ref{tab:location1}. The range of the value for the uniform prior is listed in \ref{tab:prior}. 
%
\begin{table}
    \centering
    \caption{Continuum spline points in rest-wavelength space.}
    \label{tab:location2}
    \begin{tabular}{cc}
        \hline
        Spline Point & Wavelength $\lambda_c$ (\AA) \\
        \hline
        0 & 1075 \\
        1 & 1082 \\
        2 & 1088 \\
        3 & 1095 \\
        4 & 1101 \\
        5 & 1108 \\
        6 & 1114 \\
        7 & 1121 \\
        8 & 1127 \\
        9 & 1134 \\
        10 & 1140 \\
        11 & 1147 \\
        \hline
    \end{tabular}
\end{table}
\vspace{5mm}

\begin{table}
    \centering
    \caption{Ly$\alpha$ opacity spline points in redshift space (see \href{https://github.com/JianiDing/sawtooth_project/tree/master/final_results}{final$\_$results} for details). }
    \label{tab:location1}
    \begin{tabular}{ccccc}
        \hline
        & Redshift 
        & Mean Value 
        & hpd 2.5
        &hpd 97.5 \\
        \hline
        Spline Point 12 & 2.50 & 0.27 & 0.25 & 0.29\\
        Spline Point 13 & 2.64 & 0.32 &0.31 & 0.34\\
        Spline Point 14 & 2.80 & 0.38 & 0.36& 0.39\\
        Spline Point 15 & 2.95 &0.40 &  0.38 & 0.42\\
        Spline Point 16 & 3.11 &0.46& 0.44& 0.47\\
        Spline Point 17 & 3.27 &0.52 & 0.5 & 0.54\\
        Spline Point 18 & 3.42 &0.58 & 0.56& 0.60\\
        Spline Point 19 & 3.63 &0.68  &0.67 & 0.70\\
        Spline Point 20 & 3.94 &0.85 & 0.83 & 0.87\\
        Spline Point 21 & 4.20 &1.09 & 1.07& 1.12 \\
        \hline
    \end{tabular}
\end{table}

\begin{table}
    \centering
    \caption{Ranges of the prior uniform distributions for the 22 spline points used in our MCMC inference.}
    \label{tab:prior}
    \begin{tabular}{cc}
        \hline
        Parameter & Range \\
        \hline
        Continuum spline points 0-11 & 0.6--2.8 \\
        Spline Point 12 & 0.2--0.8 \\
        Spline Point 13 & 0.245--0.845 \\
        Spline Point 14 & 0.29--0.89 \\
        Spline Point 15 & 0.335--0.935 \\
        Spline Point 16 & 0.38--0.98 \\
        Spline Point 17 & 0.47--1.07 \\
        Spline Point 18 & 0.515--1.115 \\
        Spline Point 19 & 0.56--1.16 \\
        Spline Point 20 & 0.605--1.205 \\
        Spline Point 21 & 0.65--1.25 \\
        \hline
    \end{tabular}
\end{table}



\section{Covariance Array} 

\begin{figure*}
\centering
\includegraphics[width=0.8\textwidth]{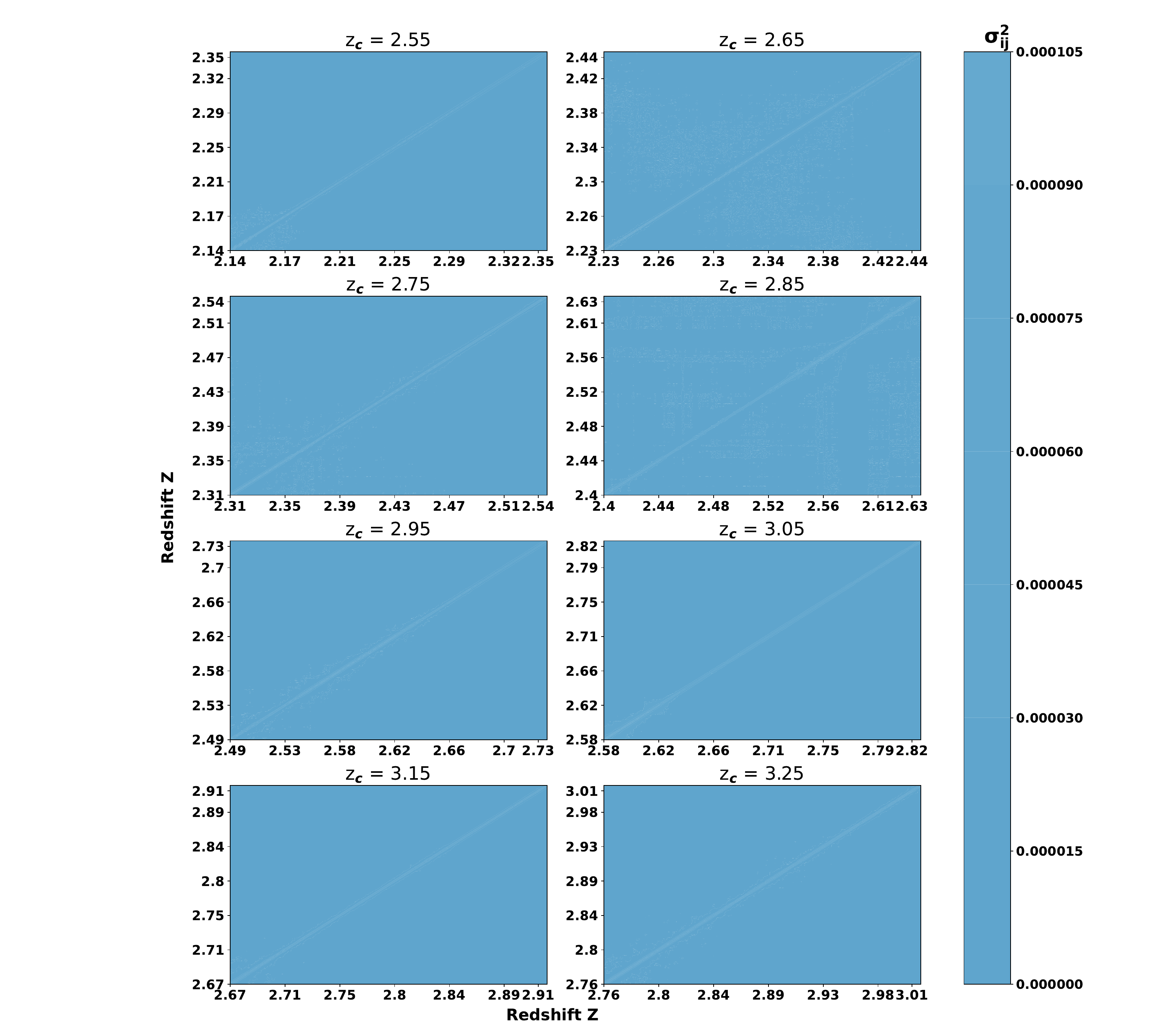}
\caption{The Covariance matrices for DR14 composite spectra in the 
redshift interval $z_c=2.85-3.55$.  All the covariance plot show similar features (sysmetric along the diagonal axis). The covariance in the off-diagonal terms are more significant in the higher redshift bins.
}
\label{fig:cov_low}
\end{figure*}

The uncertainty in the composite spectra does not follow standard error propagation because of correlations in the underlying continuum of each quasar. To assess the uncertainty and these correlations,
we generated a covariance matrix with standard bootstrap
resampling techniques.
In each iteration, a new composite was generated 
from a number $\cal N$ of randomly drawn individual spectra equal to the total number $\cal N$ of spectra in that redshift bin, allowing for duplication. This process was repeated 10,000 times to generate the covariance matrix of that composite. The covariance matrices showing correlations in the error budget of the transmitted flux are plotted 
in Figures \ref{fig:cov_low}. The structure of the covariance matrix is symmetric, and 
the variance of the transmitted flux (diagonal elements) dominates over the covariance between all possible flux pairs (off-diagonal terms). The covariance of the off-diagonal terms also increase with respect to the redshift. 

\begin{figure*}
\centering

\includegraphics[width=0.95\textwidth]{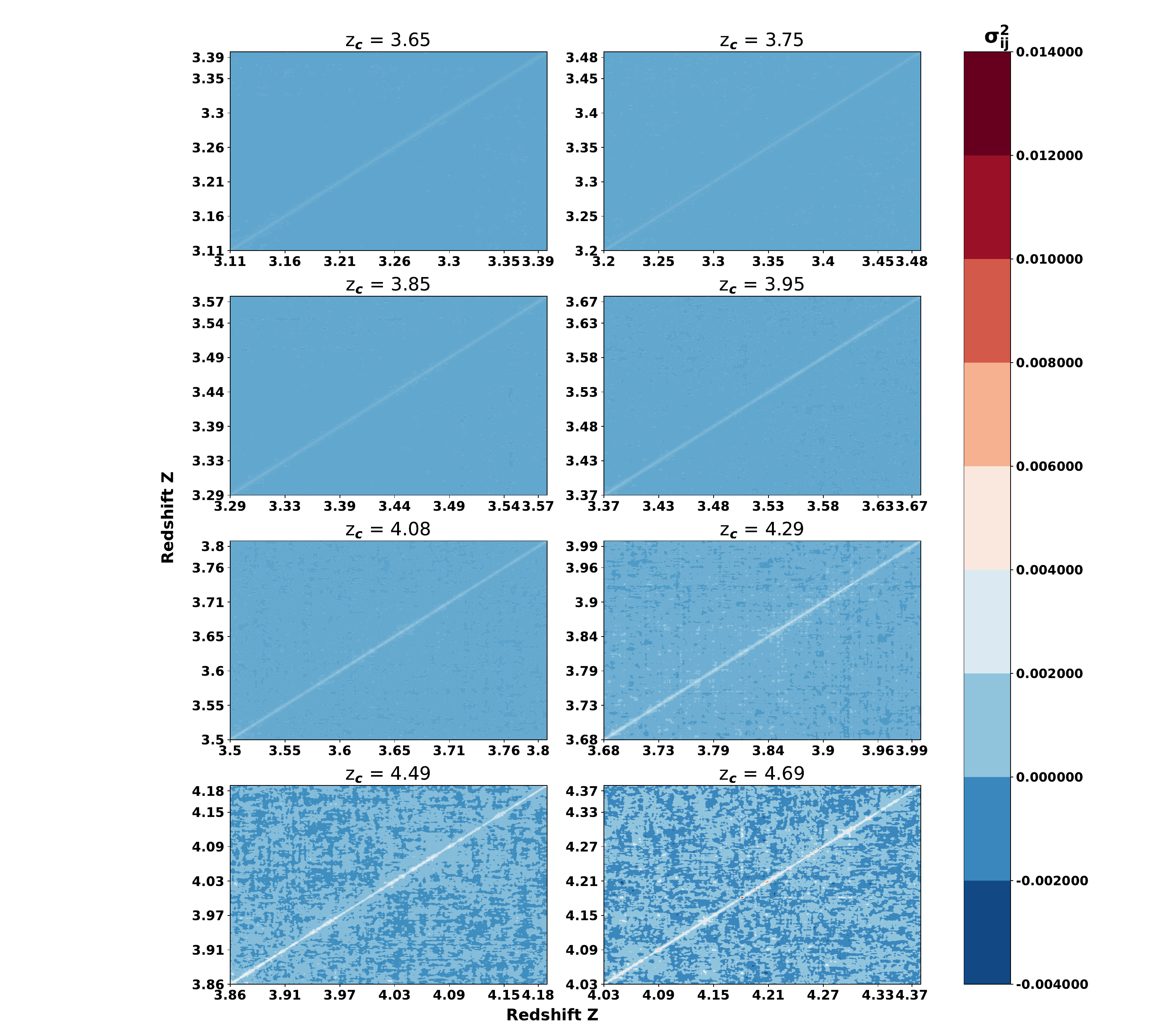}
\caption{Covariance matrices for DR14 composite spectra at $z\ge 3.65$. 
}
\label{fig:cov_high}
\end{figure*}

\section{Convergence of MCMC Algorithm}

\begin{figure*}
\centering
\includegraphics[width=160mm]{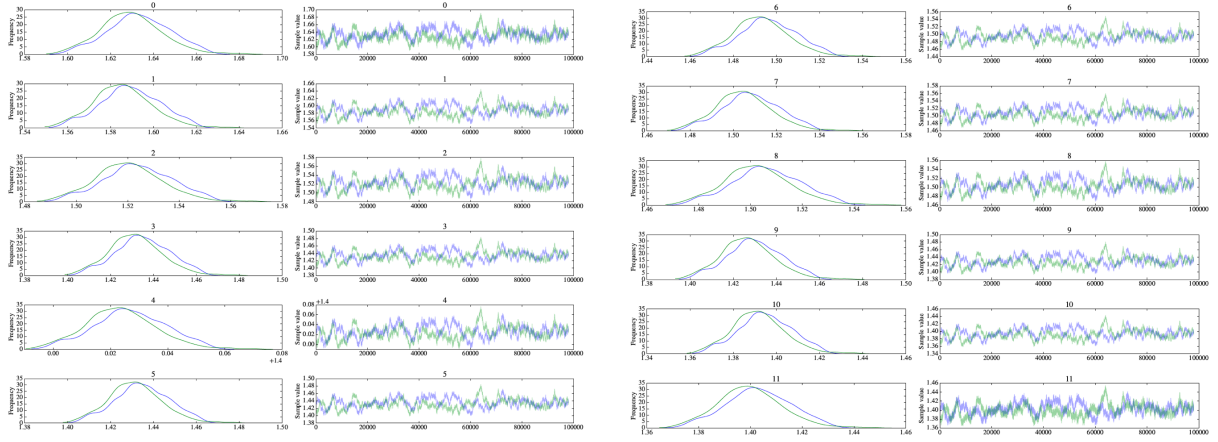}
\caption{MCMC final converged traces and kernel densities for the spline points used to represent the quasar continuum SED. The $x$-axis shows normalized flux value (left columns in each panel) and the number of the MCMC samples (right columns in each panel). The left panel depicts the smoothed histogram (using kernel density estimation) of the posteriors (left columns) and the samples of the Markov chain (right columns) of the 0-5 spline points and the right panel is same plot for the 6-11 spline points. 
}
\label{tab:taualpha1}
\end{figure*}

\begin{figure*}
\centering
\includegraphics[width=180mm]{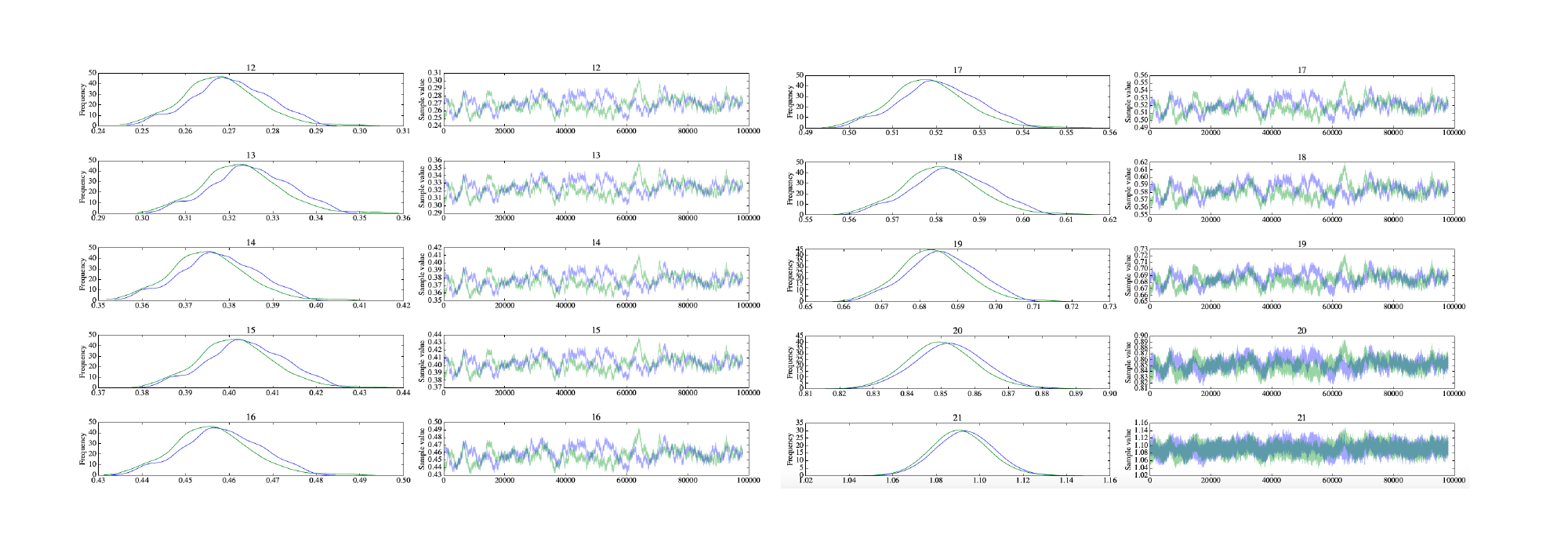}
\caption{MCMC final converged traces and kernel densities for the spline points used to represent the \lya\ opacity vs. redshift. The $x$-axis shows the value of 
\tlya\ (left columns in each panel) and the number of the MCMC samples (right columns in each panel). The left panel depicts the smoothed histogram (using kernel density estimation) of the posteriors (left columns) and the samples of the Markov chain (right columns) for the first five spline points describing the function $\tlya(z)$ (12-16 traces), and the right panel shows the next five spline points (17-21 traces).
}
\label{tab:taualpha2}
\end{figure*}

All parameters appear to have converged at the end of the MCMC sampling. Figures \ref{tab:taualpha1} and \ref{tab:taualpha2} show kernel density estimation of the posteriors and the samples of the Markov chain for all 22 spline parameters describing the quasar continuum and \tlya.

\begin{figure*}
\centering
\includegraphics[width=\textwidth]{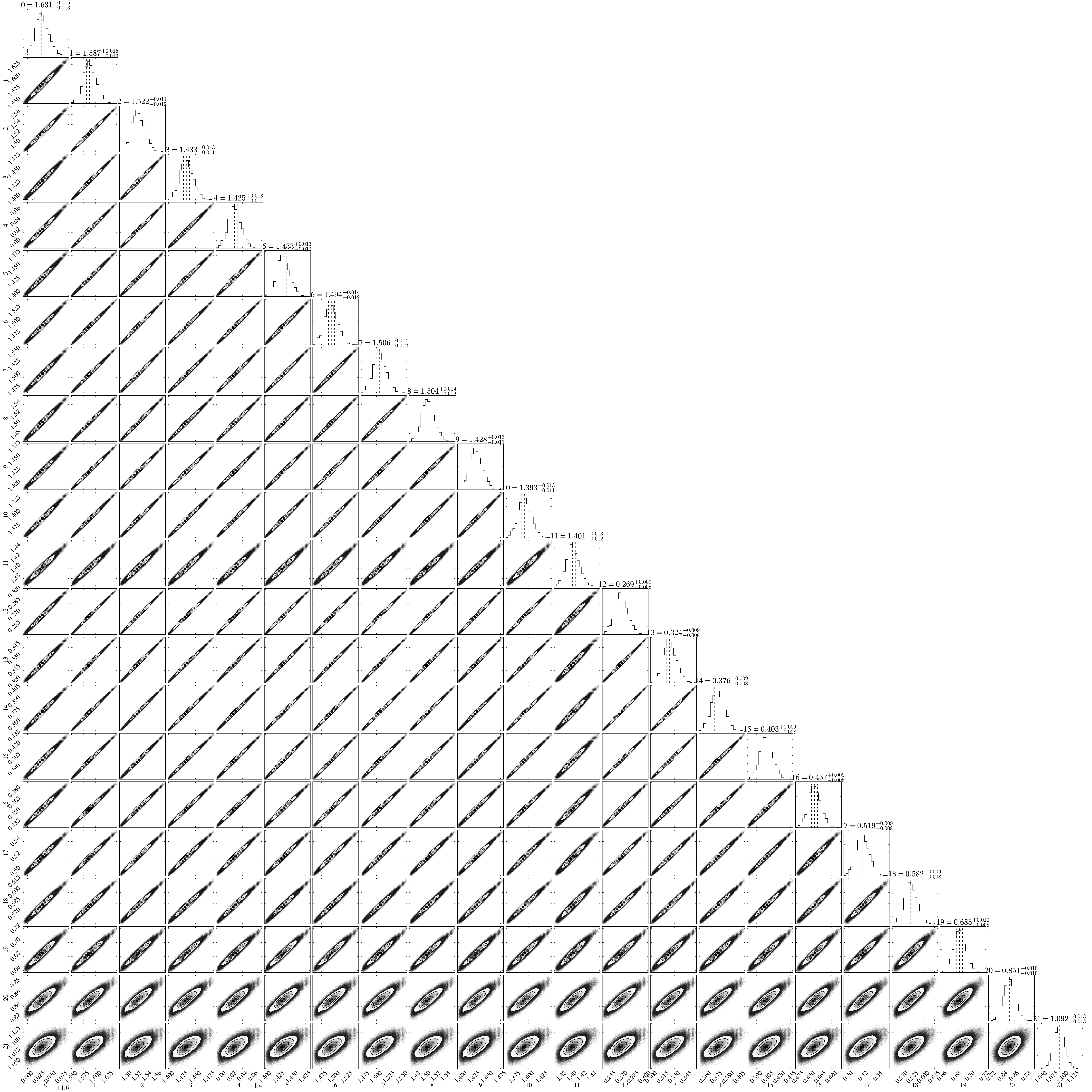}
\caption{Final corner plot for all the parameters we used to represent the continuum and $\tlya(z)$ in the MCMC fitting. 
}
\label{fig:cornerplot1}

\end{figure*}
\section{Systematic Errors Estimation from quasar SEDs}
\begin{figure*}
\includegraphics[width=0.5\textwidth]{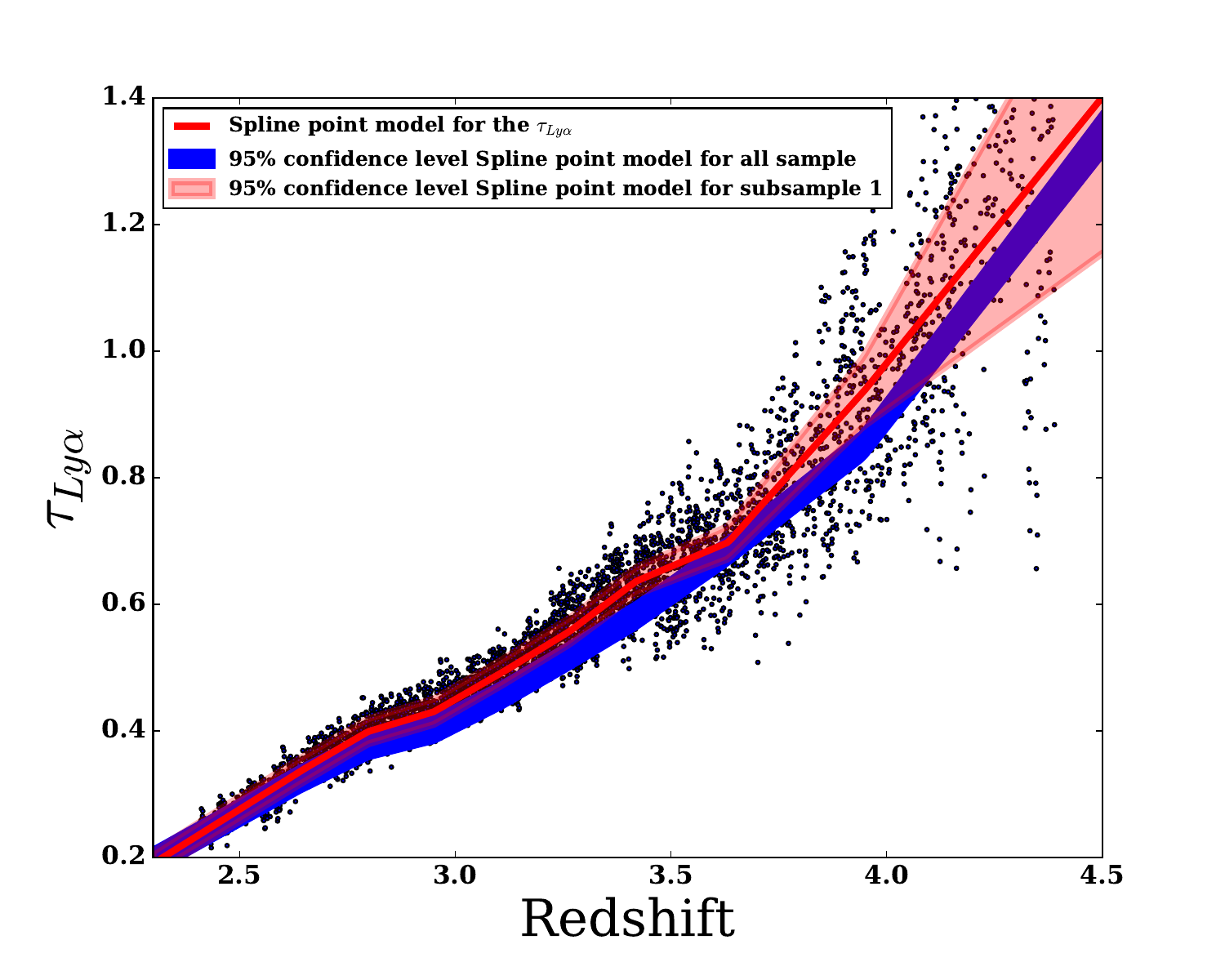}
\includegraphics[width=0.5\textwidth]{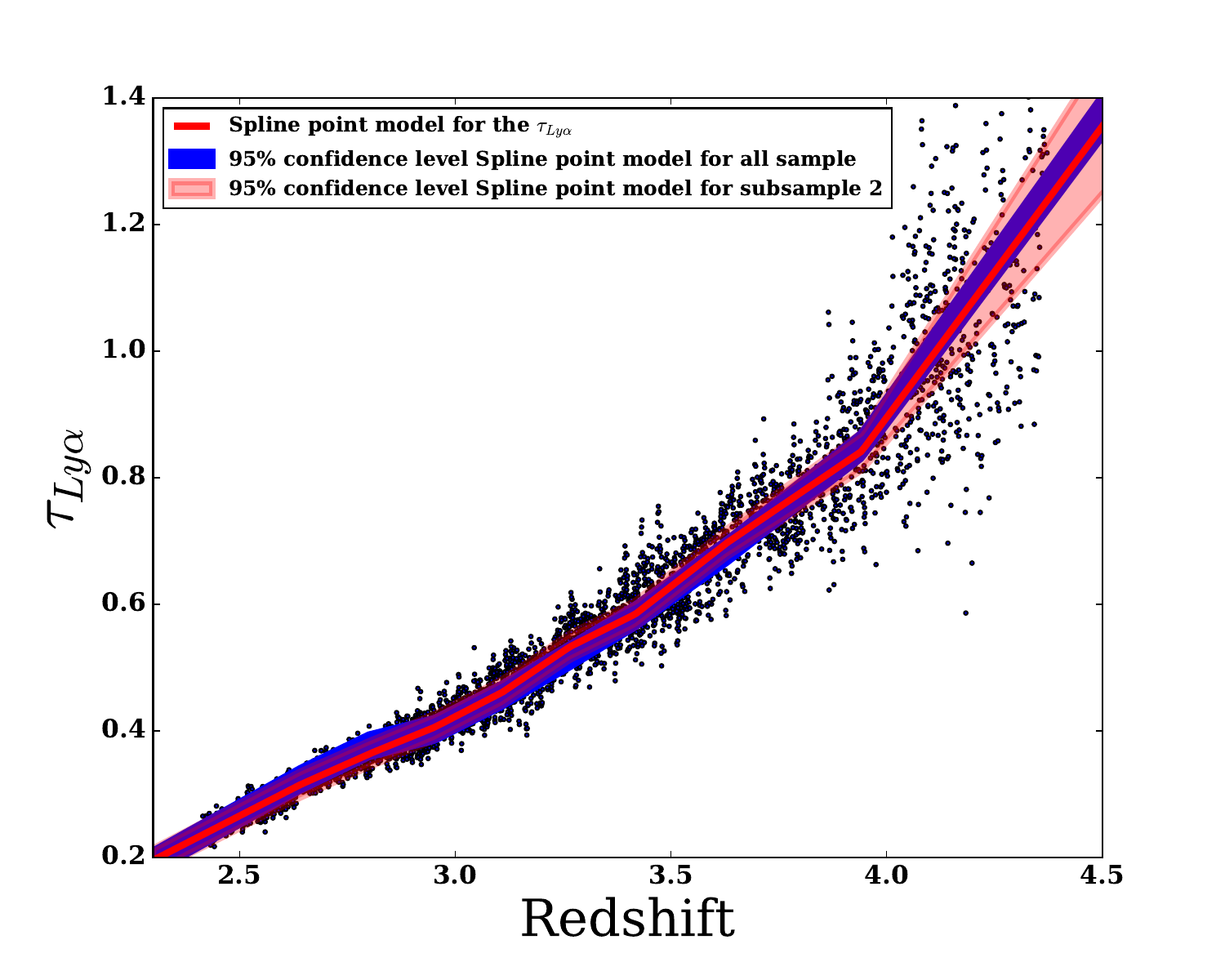}
\includegraphics[width=0.5\textwidth]{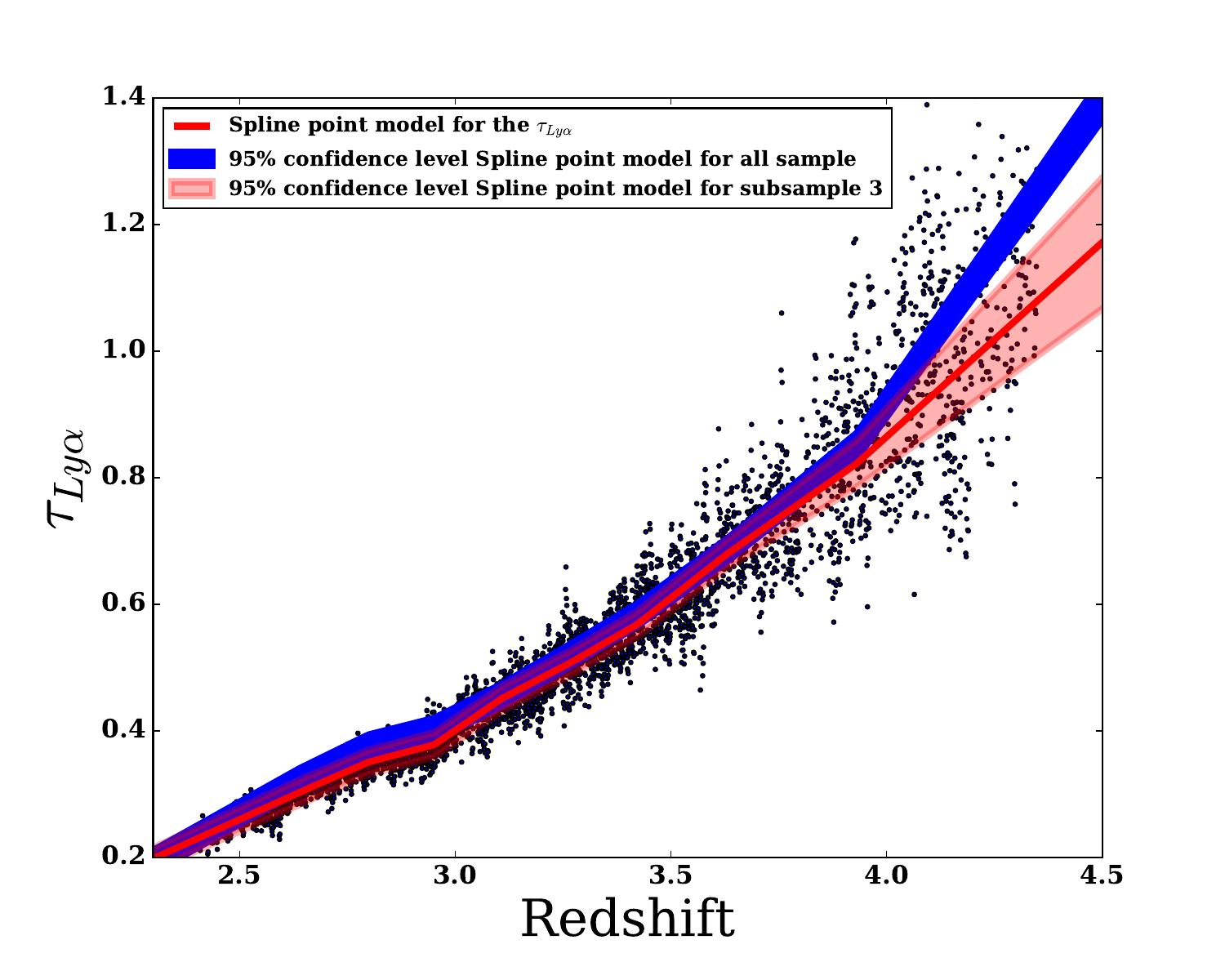}
\vspace{0cm}
\caption{\tlya\ measurement for quasar spectra with three different spectral index (-2.78$<\beta<$-2.12, -2.12$<\beta<$-1.46,-1.46$<\beta<$-0.8). The upper panel shows the comparison between the \tlya\ measurement for quasar spectra with -2.78$<\beta<$-2.12 and the \tlya\ measured with all the spectra in the sample. The 95$\%$ confidence level of the \tlya\ measurement for quasar spectra with -2.78$<\beta<$-2.12
is labeled in red and the 95$\%$ confidence level of the \tlya\ measured from all sample is labeled in blue region. The DR14 quasar data with -2.78$<\beta<$-2.12 is labeled in dark data points. The middle and lower panels are similar figures for quasar spectra with -2.12$<\beta<$-1.46 and -1.46$<\beta<$-0.8.}
\label{fig:spectralindex}
\end{figure*}

We also perform an estimation of systematic errors in the \tlya\ measurement caused by different quasar SED shapes. 
We first fit a power law $f=\alpha \lambda^{\beta}$ to the shape of the quasar SEDs, where $f$ is the flux of the quasar spectra
and $\beta$ is the spectral index. 
We then group those quasar spectra by spectral index as in \cite{kamble2020}. We adopt three spectral index bins:
$-2.78<\beta<-2.12, -2.12<\beta<-1.46, -1.46<\beta<-0.8$. We then recompute the composite spectra at all redshifts for the 
three different spectral index bins and rerun the MCMC process with our spline point method. 
The measurements of \tlya\ for the 3 bins are presented
in Figure~\ref{fig:spectralindex}.
From the differences between these \tlya\ measurements and
those from the full sample,
we estimate that the uncertainty from the systematics caused by the difference in the spectral index of the quasar SEDs is 
$6-8\%$ at $z < 4$ and $10-12\%$ at $z > 4.0$.

\clearpage

\end{document}